\documentstyle[aps,psfig,preprint]{revtex}
\draft

\begin{document}
\title{Renormalization of the weak hadronic current in the nuclear
  medium}
\author{T. Siiskonen\footnote{Contact address: EP Division, CERN, 
  CH-1211 Geneva 23, Switzerland.}\\
  Helsinki Institute of Physics, University of Helsinki, P.O. Box 9,\\
  FIN-00014 Helsinki, Finland\\ 
  M. Hjorth-Jensen\\
  Department of Physics, University of Oslo, N-0316 Oslo, Norway\\
  J. Suhonen\\
  Department of Physics, University of Jyv\"{a}skyl\"{a},
  P.O. Box 35,\\ FIN-40351 Jyv\"{a}skyl\"{a}, Finland}
\maketitle

\begin{abstract} 
The renormalization of the weak charge-changing hadronic
current as a function of the reaction energy release is studied at 
the nucleonic level.
We have calculated the average quenching factors for each type of current
(vector, axial vector and induced pseudoscalar). The obtained quenching
in the axial vector part is, at zero momentum transfer, 19\% for the $sd$ 
shell and 23\% in the $fp$ shell.
We have extended the calculations also to heavier systems such as 
$^{56}$Ni and  $^{100}$Sn,
where we obtain stronger quenchings, 44\% and 59\%, respectively.
Gamow--Teller type transitions are discussed, along with
the higher order matrix elements. The quenching factors are constant up to
roughly 60 MeV momentum transfer. Therefore the use of energy-independent
quenching factors in beta decay is justified. We also found that going 
beyond the zeroth and first order operators (in inverse nucleon mass) does 
not give any substantial contribution. The extracted renormalization to
the ratio $C_P/C_A$ at $q=100$ MeV is $-3.5$\%, $-7.1$\%, $-28.6$\%,
and $+8.7$\% for mass 16, 40, 56, and 100, respectively.
\end{abstract}

PACS number(s): 21.30.Fe, 23.40.-s, 21.60.Cs, 21.10.Pc

\section{Introduction}

The phenomenological structure of the weak hadronic current between the
proton and neutron states is well determined by its properties under the
Lorentz transformation. The additional constraints come from the
requirement of time reversal symmetry as well as from the invariance
under the $G$-parity transformation (combined charge conjugation and
isospin rotation). The resulting interaction Hamiltonian consists of
vector (V), axial vector (A), induced weak magnetism (M), and induced 
pseudoscalar (P) terms together with the associated form factors $C_\alpha$,
$\alpha=V$, $A$, $M$, or $P$. These form factors are called as coupling constants 
at zero momentum transfer. The present experimental knowledge does
not exclude the presence of the scalar and tensor interactions.
However, their contribution is expected to be small due to weak couplings 
\cite{gov98}.

The values of vector, axial vector, and weak magnetism couplings are
well established by beta-decay experiments as well as by the conserved
vector current hypothesis (CVC), introduced already in the late 50's
\cite{fey58}. The magnitude of the pseudoscalar coupling is more
uncertain, although the partially-conserved axial current hypothesis (PCAC)
\cite{gol58} provides an estimate along with muon capture experiments
in hydrogen \cite{bar81,wri98}. The value of $C_P$ in the nuclear medium is not
precisely established.

In nuclear beta decay, with an energy release up to some 20 MeV, only
the vector (Fermi) and the 
axial vector (Gamow--Teller) terms are usually important. 
The induced pseudoscalar and weak magnetism parts are essentially inactive,
since their contributions are proportional to $q/M$, where $q$ is the energy
release and $M$ is the nucleon mass (in units where $\hbar = c = 1$). 
There are, however, weak nuclear processes, like muon capture, where the
energy release is much higher (in muon capture, typically $q\approx m_\mu
\approx 100$ MeV).

Summed theoretical beta decay strengths are systematically larger than 
the experimental ones. This so-called quenching of the
(allowed) decay strength is usually explained in terms of core
polarization (degrees of freedom which are left out from the model space)
and non-nucleonic degrees of freedom like isobars and meson exchange
currents \cite{tow87}. Many authors (e.g., \cite{tow87,bro87,bro88,mar96}) have
established the quenching factors for the Gamow--Teller decays and
closely related magnetic dipole ($M1$) transitions.

In \cite{sii99,sii99j} we have self-consistently constructed effective
operators for the weak hadronic current between proton and neutron
states. These operators, as explained in Sec.\ \ref{Sec:eff}, take into
account the above-mentioned core polarization effects, which are expected
to be the largest correction to the bare matrix elements \cite{cas90}. In
the present work our aim is to calculate self-consistently the
quenching for all types of operators (V, A, M, and P) for energies up to
the muon capture range. In addition to the traditional shell-model regimes,
$sd$ and $fp$ shells, we have extended our calculations to 
$^{56}_{28}$Ni and $^{100}_{\phantom{1}50}$Sn as closed-shell cores.

Typically one expects about 20\% quenching of the axial vector part in
the $sd$ shell, i.e., the calculated Gamow--Teller matrix elements $\langle
\sigma\rangle$
are to be multiplied by a factor $\sim 0.8$ \cite{bro88} when both core 
polarization and 
non-nucleonic degrees are accounted for\footnote{The Gamow--Teller strengths
$B(\mathrm{GT})\propto\langle\sigma\rangle^2$ are then multiplied by 
$(0.8)^2\approx0.6$.}. Effects of the same magnitude are 
expected in the $fp$ shell \cite{mar96}. The authors in \cite{mar96} reach 
the conclusion that the mass dependence of the quenching has saturated 
already in the $fp$ shell. However,
the major shell closures which separate the spin-orbit partners
in $^{56}$Ni ($0f_{7/2}$ and $0f_{5/2}$, $0g_{9/2}$ and $0g_{7/2}$) and 
$^{100}$Sn ($0g_{9/2}$ and $0g_{7/2}$, $0h_{11/2}$ and $0h_{9/2}$) 
model spaces introduce large first-order corrections to
the operators. Thus,  the situation is not analogous to the one seen in
light nuclei with closed $LS$ shells. For a recent work in the Sn mass region, 
see e.g., Ref.~\cite{kar98}. We remind the reader that the quenching is always related
to the choice of the model space.

The nuclear muon capture process can be used to extract the ratio $C_P/C_A$.
Unfortunately, the results for partial capture rates are very sensitive to 
applied nuclear model, especially to the model space and to the residual 
two-body interaction (see, e.g., \cite{sii98} and references therein). The
total rates offer perhaps a more reliable source of information, 
indicating no or only small quenching for the ratio $C_P/C_A$ \cite{kol94}. 
It is of interest to see
whether this quenching can be explained in terms of effective charges for
the axial vector and pseudoscalar operators, that is, without the
complications coming from the nuclear structure calculations.

In addition to the zeroth-order Fermi and Gamow--Teller type operators, 
our set includes the first-order terms in the transition
amplitude (first order in $q/M$ as well as velocity-dependent terms).
We shall also examine the importance of the second-order terms.
We stress that the results obtained in this work can be applied quite
generally. We have used the muon as an initial bound state lepton, but the
results apply to electron capture as well and therefore to beta decay in
general (in our calculations, the muon is nothing but a heavy electron!).

In this work, after a short review of the formalism of the
semileptonic weak processes in Sec.\ \ref{Sec:ampl} and effective
operators in Sec.\ \ref{Sec:eff}, we concentrate on the results in Sec.\ 
\ref{Sec:res}. We consider four cases, with $^{16}$O, $^{40}$Ca, $^{56}$Ni, 
and $^{100}$Sn as closed-shell cores. 

\section{Invariant amplitude and single-particle operators}\label{Sec:ampl}

After the standard nonrelativistic reduction, the semileptonic charge-changing 
weak process 
\begin{equation}\label{reaction}
  \lambda_b+p\to \nu_\lambda+n,
\end{equation}
where $\lambda_b$ is a bound (anti)lepton in an atomic $1S$ orbit and
$\nu_\lambda$ is the corresponding (anti)neutrino, is described by the
amplitude
\begin{equation}\label{amplitude}
  {\cal M}^2 = \sum_{\kappa u}| M_V(\kappa,u) + M_A(\kappa,u) 
  + M_P(\kappa,u) |^2.  
\end{equation} 
We take $\lambda_b$ to be a muon, with a mass $m_\mu = 105.658$ MeV. 
The form of the effective
weak hadronic current used for Eq.\ (\ref{amplitude}) is the most
general one consistent with the expected $G$-parity symmetry and time
reversal symmetry. The functions $M_\alpha(\kappa,u)$ include the form
factors,
transition operators, and angular momentum couplings. Explicitly, the
vector part is given by 
\begin{eqnarray}\label{vector}
  M_V(\kappa,u)/C_V(q^2)&=&[0lu]S_{0u}(\kappa)\delta_{lu}-\frac{1}{M}
  [1\bar lup]S'_{1u}(-\kappa)\nonumber\\
  &+&\frac{q\sqrt3}{2M}\left\lbrace \sqrt{\frac{\bar l+1}{2\bar l+3}}[0\bar l+1 u+]
    \delta_{\bar l+1,u}+\sqrt{\frac{\bar l}{2\bar l-1}}[0\bar l-1u-]\delta_{\bar l-1,u}
  \right\rbrace S'_{1u}(-\kappa)\nonumber\\
  &+&\sqrt\frac{3}{2}\frac{q}{M}(1+\mu_p-\mu_n)\left\lbrace \sqrt{\bar l+1}\,
    W(11u\bar l,1\bar l+1)[1\bar l+1u+]\right.\\
  &+&\left. \sqrt{\bar l}\,W(11u\bar l,1\bar l-1)[1\bar l-1u-]
  \right\rbrace S'_{1u}(-\kappa). \nonumber
\end{eqnarray}
We consider the $V$ and $M$ terms together, as
suggested by the CVC: $C_M = (\mu_p-\mu_n)C_V/2M$, where $\mu_p$ and $\mu_n$ are
the anomalous magnetic moments of the proton and neutron in nuclear magnetons.
The Fermi-type term is the first term on the right hand side of Eq.\
(\ref{vector}). The axial vector part is given by
\begin{equation}\label{axial}
  M_A(\kappa,u)/C_A(q^2)=-[1lu]S_{1u}(\kappa)+\frac{1}{M}[0\bar lup]
  \delta_{\bar lu}S'_{0u}(-\kappa)-M_P(\kappa,u)/C_P(q^2),
\end{equation}
including the pseudoscalar part
\begin{equation}\label{pseudo}
  M_P(\kappa,u)/C_P(q^2)=-\frac{q}{2\sqrt3 M}\left\lbrace 
    \sqrt\frac{\bar l+1}{2\bar l+1}[1\bar 
    l+1u+]+\sqrt\frac{\bar l}{2\bar l+1}[1\bar l-1u-]\right\rbrace 
  \delta_{\bar lu}S'_{0u}(-\kappa).
\end{equation}
The Gamow--Teller-type term is the first term on the right hand side of 
Eq.\ (\ref{axial}). 

In Eqs.\ (\ref{vector})-(\ref{pseudo}), $\kappa$ labels the 
quantum numbers of the emitted neutrino $\nu_\lambda$,
\begin{eqnarray}
  \kappa &>& 0:\quad j=l-\frac{1}{2},\ l=\kappa\\
  \kappa &<& 0:\quad j=l+\frac{1}{2},\ l=-\kappa-1,
\end{eqnarray}
where $l$ and $j$ are the orbital and total angular momentum quantum numbers
of $\nu_\lambda$.
The quantity $\bar l$ is given by $l-{\rm Sign}(\kappa)$, $W$ are the
usual Racah coefficients, and 
\begin{eqnarray}
  S_{ku}(\kappa)&=&\sqrt{2(2j+1)}\,W(1/2\,1\,j\,l,1/2\,u),\quad k=1
  \nonumber\\
  &=&\sqrt{\frac{2j+1}{2l+1}},\quad k=0\\
  S'_{ku}(-\kappa)&=&{\rm Sign}(\kappa)S_{ku}(-\kappa).\nonumber
\end{eqnarray}

The most important ingredients, the transition operators, are embedded in
the reduced matrix elements $[kwu]$, $[kwu\pm]$, and $[kwup]$. The
quantum numbers labelling the matrix elements are ${\bf k}={\bf
s}_\lambda+{\bf s}_\nu$, so that $k\equiv|{\bf k}|=0$ or 1, and $u\equiv 
|{\bf u}|=|
{\bf J}_f+ {\bf J}_i|$ is the tensorial rank of the transition operator. 
The symbol $w$ is the rank of the spherical harmonics and is therefore
related to the parity change. 
It is given by $w=l$ for $[kwu]$ and
$[kwu\pm]$ type matrix elements, and $w=l+1$ or $w=l-1$ for $[kwup]$ type
matrix element ($k$ and $w$ must be able to couple to $u$). 
The symbol $p$ labels the momentum
dependent operators. The matrix elements with the corresponding
single-particle operators are listed in Table \ref{table:operators}.
These operators are further multiplied by the radial wave function of the 
initial state lepton \cite{mor60}. We have taken into account the large 
component
\begin{equation}
  G_\mu(r)=2(\alpha Zm'_\mu)^{3/2}e^{-\alpha Zm'_\mu r},
\end{equation}
where $\alpha\approx 1/137$ is the fine structure constant and $m'_\mu$ is
the reduced muon mass. 

The amplitude (\ref{amplitude}) can be used for the calculations of muon (or
electron) capture rates \cite{sii98,mor60}. As mentioned in the 
previous Section,
our aim is to calculate the effective charges (effective form factors)
for the vector, axial vector, and
pseudoscalar parts of the amplitude, so as to help to understand the
differences between calculated and experimental rates and other observables.

For the actual calculations, we divide the reduced nuclear matrix elements 
into single-particle and many-body parts,
\begin{equation}\label{obtd}
  M_\alpha(\kappa,u) = \sum_{pn}(n||m_\alpha(\kappa,u)||p)
  \frac{(J_f||[a^\dagger_n\tilde a_p]^J|| J_i)}{\sqrt{2J+1}},
\end{equation}
where $n\equiv(n_n,l_n,j_n)$ and $p\equiv(n_p,l_p,j_p)$ label the 
single-particle states. The doubly-barred
matrix elements are reduced in the angular momentum space. These reduced
single-particle matrix elements, which we calculate in the harmonic
oscillator basis with the single-particle operators $m_\alpha
(\kappa,u)$, are constructed as described 
in the following Section. Further, we have defined
\begin{equation}\label{atilde}
  \tilde a_{jm}=(-1)^{j+m}a_{j,-m}.
\end{equation}

As an example, from Eq.\ (\ref{axial}), the single-particle matrix element 
for the axial vector part is
\begin{eqnarray}\label{mp}
  (n||m_A(\kappa,u)||p)&=&-C_A(q^2)(n||O_{1lu}G_\mu(r)||p)S_{1u}(\kappa)+
  \frac{C_A(q^2)}{M}(n||O_{0\bar lup}G_\mu(r)||p)
  \delta_{\bar lu}S'_{0u}(-\kappa)\nonumber\\
  &+&C_A(q^2){q\over2\sqrt{3}M}\left\lbrace
    \sqrt{\bar l+1\over 2\bar l+1}
    (n||O_{1\bar l+1u+}G_\mu(r)||p)\right.\\
  &+&\left.\sqrt{\bar l\over 2\bar l+1}(n||
    O_{1\bar l-1u-}G_\mu(r)||p)\right\rbrace\delta_{\bar lu}S'_{0u}
  (-\kappa).\nonumber
\end{eqnarray}
A closer look into the Gamow--Teller type matrix element gives
\begin{eqnarray}
  -C_A(q^2)(n||O_{1lu}G_\mu(r)||p)S_{1u}(\kappa) &=& -C_A(q^2)
  i^{l_p-l_n}(l_nj_n||Y_{1lu}(\hat{\bf r},\sigma)||l_pj_p)\nonumber\\
  &\times&\int\limits_0^\infty r^2 R_{n_nl_n}(r)
  j_l(qr)R_{n_pl_p}G_\mu(r)dr
  \sqrt{2(2j+1)}\,W(1/2\,1\,j\,l,1/2\,u).
\end{eqnarray}
Here $R_{nl}(r)$ are the radial single-particle wave functions. The reduced
matrix element of the vector spherical harmonics is given by
\begin{eqnarray}
  (l_nj_n||Y_{1lu}(\hat{\bf r},\sigma)||l_pj_p)&=&
  \sqrt{3\over 16\pi^2}(-1)^{l_p+j_p+j_n+l+1}\hat l
  \hat u\hat j_p\hat j_n\left(\begin{array}{ccc} j_p & j_n & u \\ 1/2 & 
      -1/2 & 0
    \end{array}\right)\\
  &\times&\left[{\hat j_n^2+(-1)^{j_n+j_p+u}\hat j_p^2\over \sqrt{2u(u+1)}}
    \left(\begin{array}{ccc} u & 1 & l \\ 1 & -1 & 0 \end{array}\right)
    +(-1)^{l_n+1/2+j_n}\left(\begin{array}{ccc} u & 1 & l \\ 0 & 0 & 0 
      \end{array}
    \right)\right],\nonumber
\end{eqnarray}
where $\hat x = \sqrt{2x+1}$.
The expressions for the vector and pseudoscalar parts are obtained in a 
similar way
using Eqs.\ (\ref{vector}) and (\ref{pseudo}) and Table \ref{table:operators},
where the operators $O$ of Eq.\ (\ref{mp}) are given. For more details, see
\cite{mor60}. 

The many-body part, which includes 
the one-body transition density (OBTD), is given by
the adopted nuclear model. The effective operators introduced in Sec.\
\ref{Sec:eff} do not affect the OBTD, which are taken to be given
numbers. We do not calculate them here (see, e.g., \cite{sii98} 
for examples). In what follows we shall consider only the single-particle
part of Eq.\ (\ref{obtd}). This is analogous to, e.g., $M1$-transitions, where
the effective charges are calculated without referring to the many-body
part. 

\section{Perturbative methods and effective operators to second order}
\label{Sec:eff}

In order to derive a microscopic approach to the effective operator
within the framework of perturbation theory, we need to introduce various
notations and definitions pertinent to the methods exposed.
In this section we briefly review how to calculate an effective one-body
operator within the framework of degenerate Rayleigh-Schr\"{o}dinger
(RS) perturbation theory \cite{ko90,lm85}, see also Refs.~\cite{tow87,cas90}
for a detailed discussion on various effective operator diagrams.

It is common practice in perturbation theory to reduce the infinitely
many degrees of freedom of the Hilbert space to those represented
by a physically motivated subspace, the model space.
In such truncations of the Hilbert space, the notions of a projection
operator $P$ onto the model space and its complement $Q$ are
introduced. The projection operators defining the model and excluded
spaces are defined by
\begin{equation}
  P=\sum_{i=1}^{D} \left|\Phi_i\right\rangle
  \left\langle\Phi_i\right |,
\end{equation}
and
\begin{equation}
  Q=\sum_{i=D+1}^{\infty} \left|\Phi_i\right\rangle
  \left\langle\Phi_i\right |,
\end{equation}
with $D$ being the dimension of the model space, and $PQ=0$, $P^2 =P$,
$Q^2 =Q$, and $P+Q=I$. The wave functions $\left|\Phi_i\right\rangle$ 
are eigenfunctions
of the unperturbed hamiltonian $H_0 = T+U$, where $T$ is the kinetic
energy and $U$ an appropriately chosen one-body potential, that of the
harmonic oscillator (h.o.) in this calculation. The full Hamiltonian
is then rewritten as $H=H_0 +H_1$ with $H_1=V-U$, $V$ being e.g., the
nucleon-nucleon (NN) interaction or the $G$-matrix to be discussed below. 
The eigenvalues and eigenfunctions of the full Hamiltonian are denoted by
$\left|\Psi_{\alpha}\right\rangle$ and $E_{\alpha}$, i.e.,
\begin{equation}
  H\left|\Psi_{\alpha}\right\rangle= 
  E_{\alpha}\left|\Psi_{\alpha}\right\rangle.
\end{equation}
Rather than solving the full Schr\"{o}dinger equation above, one defines
an effective Hamiltonian acting within the model space such that
\begin{equation}
  PH_{\mathrm{eff}}P\left|\Psi_{\alpha}\right\rangle=
  E_{\alpha}P\left|\Psi_{\alpha}\right\rangle=
  E_{\alpha}\left|\Phi_{\alpha}\right\rangle
\end{equation}
where $\left|\Phi_{\alpha}\right\rangle=P\left|\Psi_{\alpha}\right\rangle$
is the projection of the full wave function
onto the model space, the model space wave function.
In RS perturbation theory, the effective interaction
$H_{\mathrm{eff}}$ can be written out order by order in the 
interaction $H_1$ as
\begin{equation}
  PH_{\mathrm{eff}}P=PH_1P +PH_1\frac{Q}{e}H_1 P+
  PH_1\frac{Q}{e}H_1 \frac{Q}{e}H_1 P+\dots.
  \label{eq:effint}
\end{equation}
Here we have defined $e=\omega -H_0$, 
where $\omega$ is the so-called starting energy, defined as the unperturbed
energy of the interacting particles. Similarly, the exact wave
function $\left|\Psi_{\alpha}\right\rangle$
can now be written in terms of the model space wave function as
\begin{equation}
  \left|\Psi_{\alpha}\right\rangle=
  \left|\Phi_{\alpha}\right\rangle+
  \frac{Q}{e}H_1\left|\Phi_{\alpha}\right\rangle
  +\frac{Q}{e}H_1\frac{Q}{e}H_1\left|\Phi_{\alpha}\right\rangle+\dots
  \label{eq:wavef}
\end{equation}

In studies of nuclear transitions such as beta decay, the quantity of
interest is the transition matrix element between an initial state
$\left|\Psi_i\right\rangle$ and a final state $\left|\Psi_f\right\rangle$
of an operator ${\cal O}$ defined as
\begin{equation}
  {\cal O}_{fi}=
  \frac{\left\langle\Psi_f\right|
    {\cal O}\left|\Psi_i\right\rangle }
  {\sqrt{\left\langle\Psi_f | \Psi_f \right\rangle
      \left\langle \Psi_i | \Psi_i \right\rangle}}.
  \label{eq:effop1}
\end{equation}
Since we perform our calculation in a reduced space, the exact
wave functions $|\Psi_{f,i}\rangle$ are not known, only their
projections onto the model space. We are then confronted with the
problem of how to evaluate ${\cal O}_{fi}$ when only the model
space wave functions are known. In treating this problem, it is usual
to introduce an effective operator
${\cal O}_{fi}^{\mathrm{eff}}$, defined by requiring
\begin{equation}
  {\cal O}_{fi}=\left\langle\Phi_f\right |{\cal O}_{\mathrm{eff}}
  \left|\Phi_i\right\rangle.
\end{equation}
Observe that ${\cal O}_{\mathrm{eff}}$
is different from the original operator ${\cal O}_{fi}$. The standard
empirical procedure is then to introduce some adjustable parameters
in ${\cal O}_{fi}^{\mathrm{eff}}$. 

The perturbative expansion for the effective operator can then
be written in a much similar way as Eqs.\ (\ref{eq:effint}) and
(\ref{eq:wavef}), i.e.,  
\begin{equation}
  \left\langle\Psi_{f}\right|{\cal O}\left|\Psi_{i}\right\rangle=
  \left\langle\Phi_{f}\right|{\cal O}\left|\Phi_{i}\right\rangle+
  \left\langle\Phi_{f}\right|{\cal O}\frac{Q}{e}H_1\left|\Phi_{i}\right\rangle+
  \left\langle\Phi_{f}\right|\frac{Q}{e}H_1{\cal O}\left|\Phi_{i}\right\rangle+
  \left\langle\Phi_{f}\right|{\cal O}\frac{Q}{e}H_1\frac{Q}{e}H_1\left|\Phi_{i}
  \right\rangle+\dots
  \label{eq:effoperexp}
\end{equation}

\begin{figure}[hbtp]
{\centering
\mbox{\psfig{figure=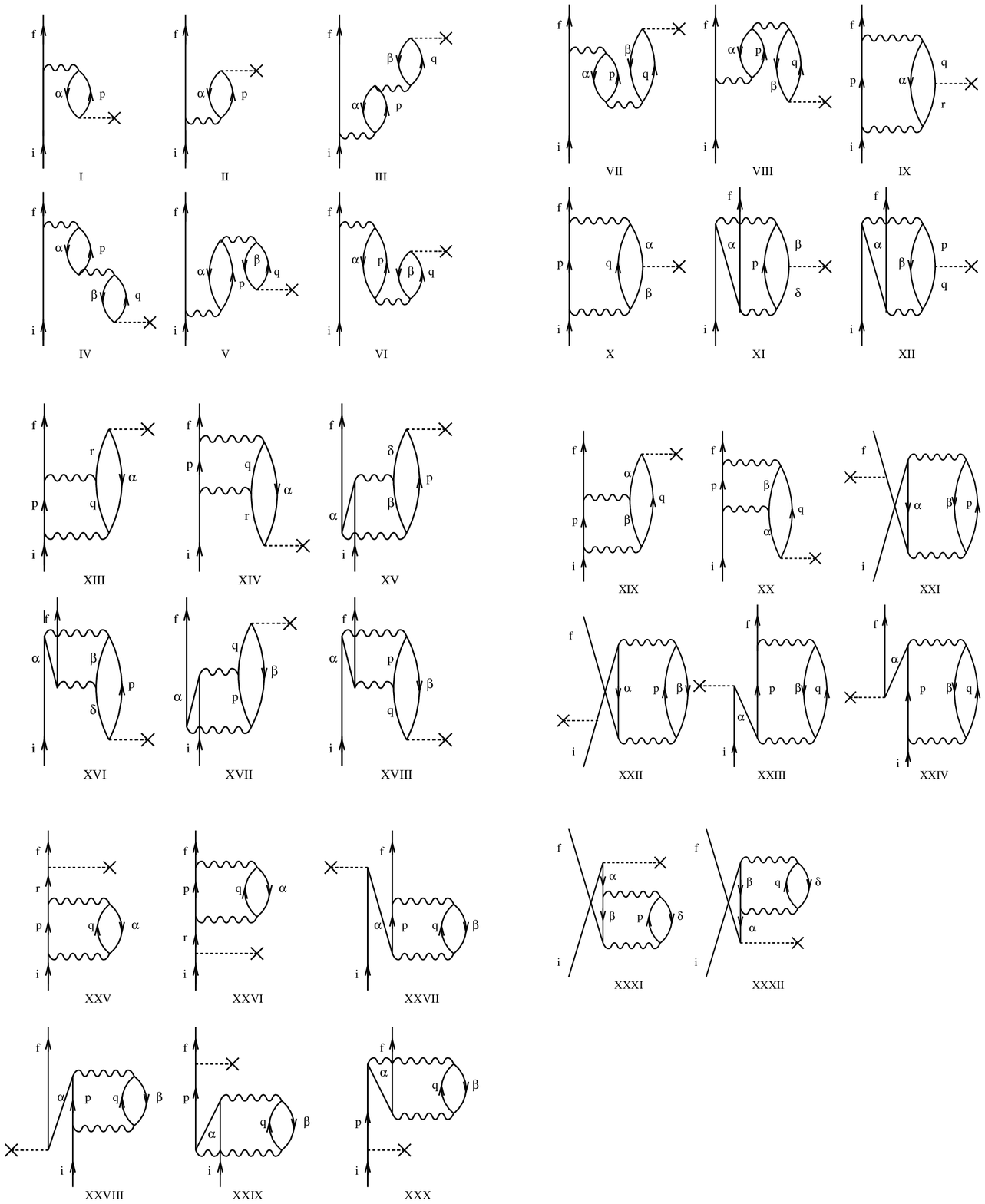,height=25cm,width=18cm,angle=0}}
}
       \caption{Non-folded diagrams to second order in the interaction 
                included in the evaluation of the effective 
                operator. Hole states are represented by greek letters while
                particle states are given by roman letters. The operator 
               itself is given by $---\times$ in the various diagrams, 
               while the wiggly lines are the
               nuclear $G$-matrix. Folded diagrams to second order 
               in the interaction are included in the calculation but 
               not shown here. }
       \label{fig:effop2nd}
\end{figure}

In Fig.\ \ref{fig:effop2nd} we list all diagrams (except folded diagrams)  
to second order in the interaction evaluated in this
work. We do not include Hartree-Fock insertions. For pure  
Gamow--Teller or Fermi like operators,
see e.g., the review article by Towner \cite{tow87}, such diagrams are exactly
zero. 
Another feature of e.g., the Gamow--Teller-type operators  
is that for several diagrams involving particle-hole contributions,
these diagrams are exactly zero unless the particle-hole orbits are
spin-orbit partners. This means that for $LS$-closed shell nuclei 
like $^{16}$O and $^{40}$Ca diagrams like (I)-(VIII) or 
(XIII)-(XX) are all zero. However, this picture changes when we
move to closed-shell nuclei like 
$^{56}$Ni and $^{100}$Sn. For Ni the last proton and neutron holes
are in the $0f_{7/2}$ single-particle orbit. This means that the  
$0f_{7/2}$ hole and the $0f_{5/2}$  
particle states in $^{56}$Ni yield non-vanishing contributions to 
the Gamow--Teller-type operator from the above-mentioned diagrams. Similarly, 
in Sn these contributions are represented by the spin-orbit partners 
in the the $0g_{9/2}$  hole and the $0g_{7/2}$ 
particle states. These spin-orbit partners yield then 
$1\hbar\omega$ intermediates states. Similarly, we have also 
spin-orbit partners for particles states outside the model space.
These are $0g_{9/2}$ and  $0g_{7/2}$ for $^{56}$Ni and 
$0h_{11/2}$ and  $0h_{9/2}$
for $^{100}$Sn. These $1\hbar\omega$ intermediates states are 
then responsible for the different quenching 
of the effective operators in the mass regions of 
$^{16}$O-$^{40}$Ca and $^{56}$Ni-$^{100}$Sn, respectively. 

We end this section with a discussion of how to construct a 
$G$-matrix. The $G$-matrix enters in turn our perturbative expansion
for the effective operator.
As is well know in nuclear physics, the NN potential
exhibits a repulsive core, which renders any perturbative
treatment prohibitive. However, one possible way of overcoming
this deficiency is to introduce
the reaction matrix $G$, which accounts for
short range correlations. The $G$-matrix is defined through
\begin{equation}
  G=V+V\frac{Q}{\omega - QTQ}G.
  \label{eq:gmat}
\end{equation}
Here, $\omega$ is the energy of the interacting nucleons in
a medium and $V$ is the free NN potential.
We have assumed that the energy of the intermediate
states can be replaced by the free kinetic spectrum $T$, since
these states are predominantly of high excitation energy.

In this work we solve Eq.\ (\ref{eq:gmat}) for finite nuclei by 
employing a formally exact technique for handling $Q$, originally 
presented by Tsai and Kuo \cite{tk72} and discussed in Ref.\ 
\cite{kkko76}. Tsai and Kuo employed the matrix identity
\begin{equation}
  Q\frac{1}{QAQ}
  Q=\frac{1}{A}-
  \frac{1}{A}P\frac{1}{P
    A^{-1}P}P\frac{1}{A},
\end{equation}
with $A=\omega -T-V$, to rewrite Eq.\ (\ref{eq:gmat}) as
\begin{equation}
  G = G_{F} +\Delta G,\label{eq:gmod}
\end{equation}
where $G_{F}$ is the free $G$-matrix defined as
\begin{equation}
  G_{F}=V+V\frac{1}{\omega - T}G_{F}. \label{eq:freeg}
\end{equation}
The term $\Delta G$ is a correction term defined entirely within the
model space $P$ and given by
\begin{equation}
  \Delta G =-V\frac{1}{A}P
  \frac{1}{PA^{-1}P}P\frac{1}{A}V.
\end{equation}
Employing the definition for the free $G$-matrix of Eq.\ (\ref{eq:freeg}),
one can rewrite the latter equation as
\begin{equation}
  \Delta G =-G_{F}\frac{1}{e}P
  \frac{1}{P(e^{-1}+e^{-1}G_{F}e^{-1})
    P}P\frac{1}{e}G_F,
\end{equation}
with $e=\omega -T$.

We see then that the $G$-matrix for finite nuclei is expressed as the sum 
of two terms; the first term is the free $G$-matrix with no Pauli corrections
included, while the second term accounts for medium modifications
due to the Pauli principle. The second term can easily be obtained by some 
simple matrix operations involving the model-space matrix $P$ only. 

Finally, in order to calculate the $G$-matrix for the various mass regions,
we need to define the relevant model spaces used to define the $P$ and
$Q$ operators in the equation for $G$. The oscillator energies  
$\hbar\Omega$ will be derived from $\hbar\Omega = 45A^{-1/3} - 25A^{-2/3}$, 
$A$ being the mass number. This yields $\hbar\Omega =13.9$,
$\hbar\Omega =11.0$, $\hbar\Omega =10.05$, and $\hbar\Omega =8.5$ 
MeV for $A=16$, $A=40$, $A=56$, and $A=100$, respectively.
We choose the model spaces which are believed, from
both experiment and theoretical calculations, to be relevant
as a first approximation for calculations of effective
interactions and operators 
in the mass areas from $A=16$ to $A=100$. These are
the $0d_{5/2}$, $0d_{3/2}$, and $1s_{1/2}$ orbits
for $A=16$, the $1p_{3/2}$, $1p_{1/2}$, $0f_{7/2}$, and $0f_{5/2}$ orbits
for nuclei in the mass region of $A=40$, the $1p_{3/2}$,
$1p_{1/2}$, $0f_{5/2}$, and $0g_{9/2}$ orbits
for nuclei in the mass region of $A=56$ and the
$0h_{11/2}$, $0g_{7/2}$, $1d_{5/2}$, $1d_{3/2}$, and $2s_{1/2}$ orbits
for $A=100$. For these systems, the closed-shell cores ($^{16}$O,
$^{40}$Ca, $^{56}$Ni, and $^{100}$Sn) have equal numbers of protons and neutrons,
and the model spaces are the same for both protons and neutrons. 

The definition of the Pauli operator for the $G$-matrix can be found
in Refs.\ \cite{kkko76,hko95}, where the so-called double-partitioned
scheme has been used. A detailed discussion
on the computation of the $G$-matrix can be found in Ref.~\cite{hko95}.
This definition means that also the shell above that which defines
the model space of the effective interaction, is included in the
evaluation of the $G$-matrix. For the $1s0d$-shell, this means that we
also include the $1p0f$ shell in the definition of the $P$-operator for 
the $G$-matrix.
As a consequence, we have to include in our perturbation expansion
ladder type of diagrams
where the allowed intermediate states are those of the $1p0f$-shell.
With this prescription, we have to evaluate  diagrams X, XIII, XIV, and
XXIII-XXVIII in Fig.\ \ref{fig:effop2nd}.

In our actual calculation of the various effective operators, we truncate
the sum over intermediate states at excitations of $4-8\hbar\omega$
in oscillator energy. This truncation yields and error of $\sim 1\%$
in our evaluation of the effective operator.  
The nucleon-nucleon interaction employed in this work is the 
CD-Bonn interaction of Machleidt {\em et al.} \cite{mac96}.

\section{Results}\label{Sec:res}

We consider the transition $0^+\to 1^+$, which is of rank
$u=1$. In addition, $\kappa=-1$ or $\kappa=2$, corresponding to
$j=\frac{1}{2}$ and $l=0$ or $j=\frac{3}{2}$ and
$l=2$, respectively. The nuclear matrix elements, allowed by these
quantum numbers together with parity conservation, are $[101]$, $[121]$,
$[101-]$, $[121+]$, $[011p]$, and $[111p]$. We remind the reader that the
matrix element $[101]$ is closely related to the Gamow--Teller matrix
element of the nuclear beta decay, only the radial dependence is more
complicated due to the possibility of a larger energy release (see Table
\ref{table:operators}). Our set includes matrix elements which are classified 
as forbidden in the nuclear beta decay \cite{mor60}.

In order to discuss the average quenching, we define a factor $\rho_\alpha$
\begin{equation}\label{rho}
  \rho_\alpha(q)\equiv\rho_\alpha=\frac{\sum_{pn}|(n||m_\alpha^{\rm
      ren}(\kappa,u)||p)|}{\sum_{pn}|(n||m_\alpha^{\rm bare}(\kappa,u)
      ||p)|},
\end{equation}
where $\alpha=V$, $A$, or $P$, and ''ren'' and ''bare'' refer to renormalized and
bare single-particle matrix elements, respectively. The sums run over all
the single-particle states included in the model space. We use absolute
values in the sums in order to avoid cases where two single-particle
matrix elements, with same magnitudes and opposite signs, cancel each
other. This kind of cancellations do not easily happen in nuclear
structure calculations, since the involved OBTD have different
magnitudes (and signs).

\begin{figure}
{\centering
\mbox{\psfig{figure=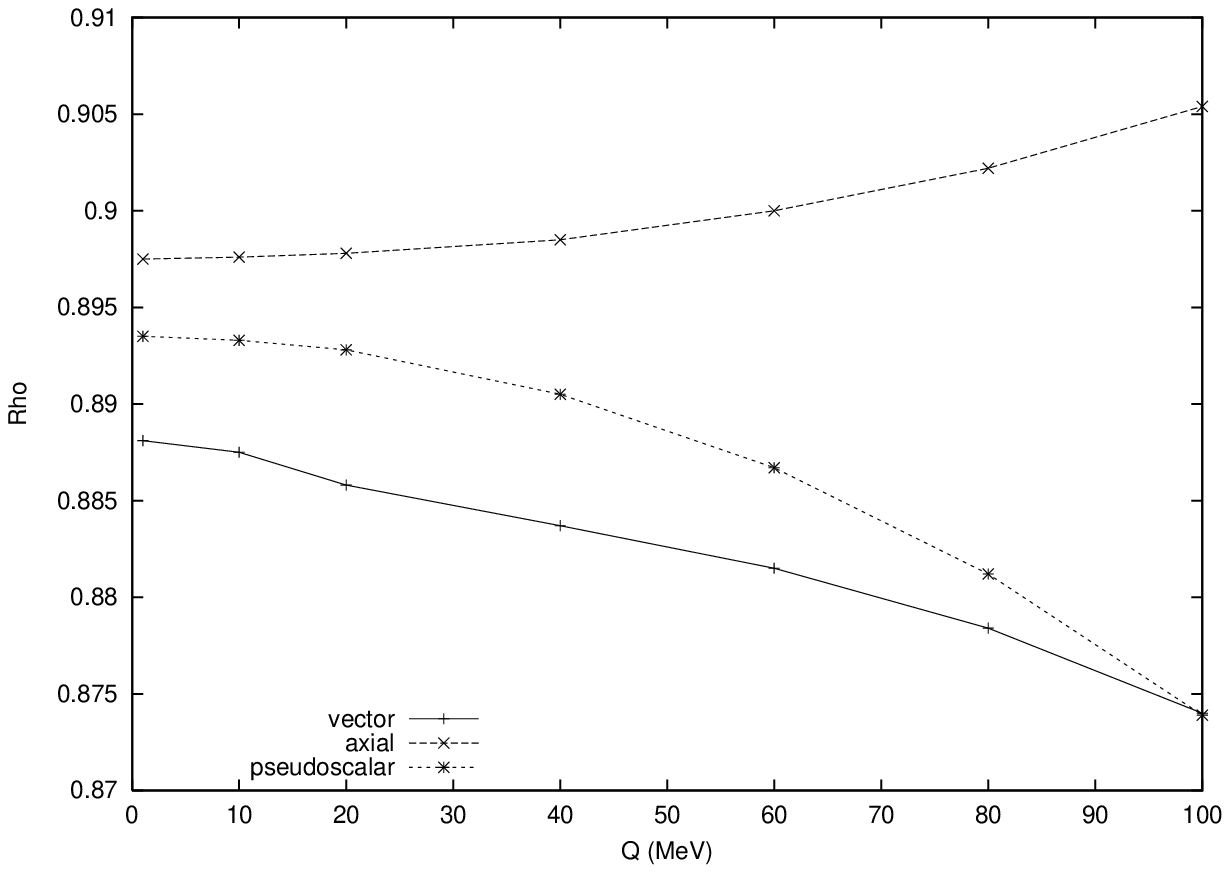,height=10cm,width=15cm,angle=0}}
}
  \caption{Renormalization of the vector, axial vector 
    and pseudoscalar terms with $^{16}$O as the closed-shell
    core, $\kappa=-1$ ($J=0\to 1$ transition, $\Delta\pi=\mathrm{no}$).
    \label{Fig:16}}
\end{figure}

We start the discussion with $sd$ shell nuclei. The effective operators 
are calculated with $^{16}_{\phantom{1}8}$O as a closed-shell core. 
The model space is the full $1s0d$ shell. From
Fig.\ \ref{Fig:16} we see that the quenching for $V$, $A$, and $P$ terms
remains essentially constant for the whole energy range considered. In
particular, at the beta decay energies (below 20 MeV) all $\kappa=-1$
terms are constant. Moreover, we have
$\rho^2_A(0)=0.81$ which is somewhat larger value than the empirical 
''universal
quenching factor'' of Ref.\ \cite{bro88}, fitted to a large body of
beta decay data in the $sd$ shell (we remind the reader that our effective
operators do not include the subnucleonic degrees of freedom). We also get a
clear renormalization in the vector and pseudoscalar parts of the current.
Note that CVC does not apply here, since we are not looking at
Fermi transitions, the vector-type contribution comes from the higher-order
terms. 

Strictly speaking, our quenching factors are applicable only for one-particle
systems, e.g., for $^{17}$O. In practice, these factors are used for the whole
model space, and the (weak) mass dependence is simply left out.
In fact, we find that $\rho_A^2(0)=0.81$ for $^{28}_{14}$Si as well 
(at the one-particle level). At the end of
the model space, effective operators derived for the hole state can also be
used. Then, often a larger renormalization compared to the particle operator
is seen \cite{tow87}, reflecting the mass dependence.

\begin{figure}
{\centering
\mbox{\psfig{figure=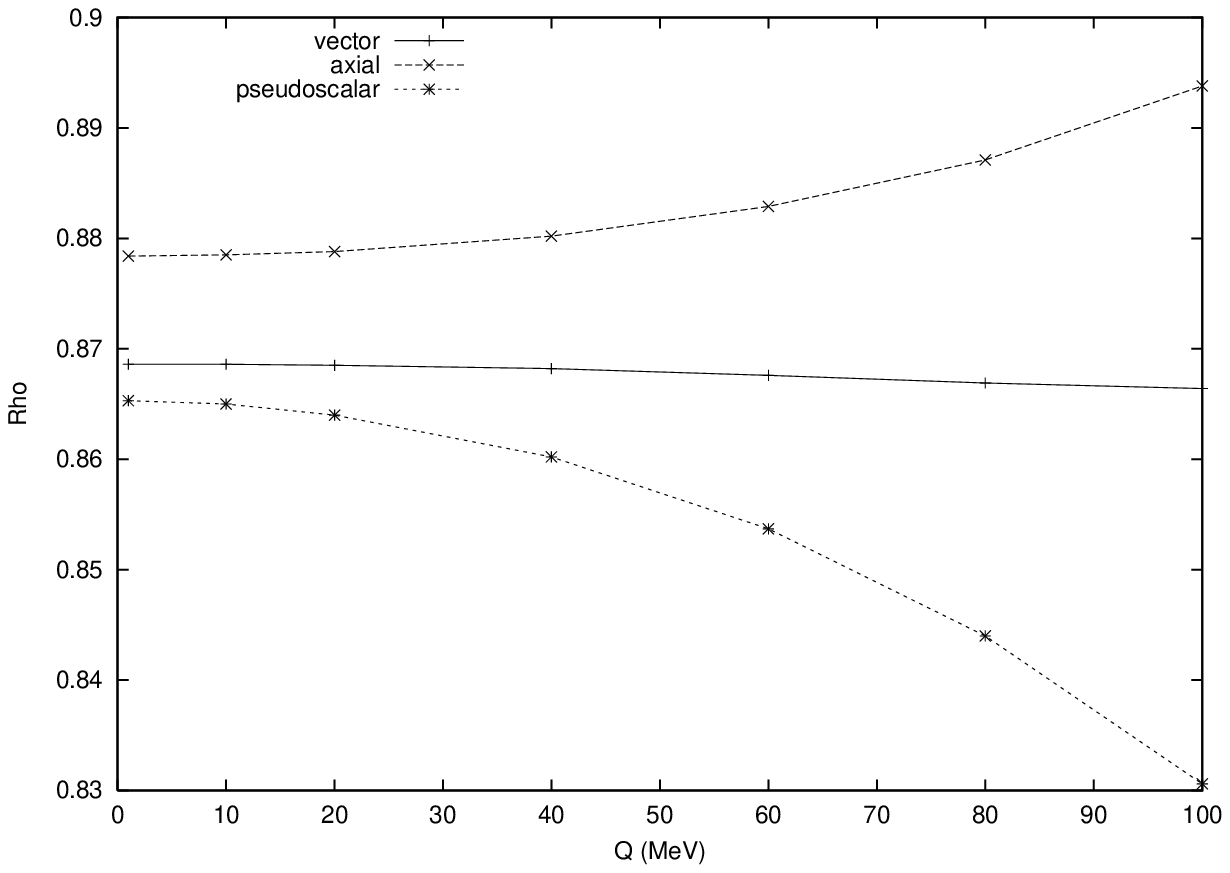,height=10cm,width=15cm,angle=0}}
}
  \caption{Renormalization of the vector, axial vector 
    and pseudoscalar terms with $^{40}$Ca as the closed-shell
    core, $\kappa=-1$ ($J=0\to 1$ transition, $\Delta\pi=\mathrm{no}$).
    \label{Fig:40}}
\end{figure}

For $^{40}_{20}$Ca as a core, Fig.\ \ref{Fig:40}, the overall features are 
very similar to $^{16}$O. We get slightly more quenching, $\rho^2_A(0)=0.77$. 
This is in line with earlier studies, where only a slightly larger
quenching for the $fp$ shell is introduced. Our model space includes all the
single-particle orbits of the $0f1p$ shell. 
In \cite{mar96} the authors reach the conclusion that already
in the $fp$ shell the quenching factor has reached the large-$A$ limit. This
is, indeed, confirmed by our results. We also note the good agreement
with the results of Ref.\ \cite{tow94}, Table 1. 

Qualitatively,
the saturation of the quenching comes from the similar choice of the
model space (complete $0\hbar\omega$ space) in $^{16}$O and $^{40}$Ca. 
In both cases, the first-order diagrams
give zero contribution. Therefore the second-order contribution, with
$1\hbar\omega$ intermediate excitations, is the most important one, 
and very similar behaviour is expected and, indeed, seen.
The differences can be attributed e.g., to different oscillator
parameters and differences in the single-particle orbit structures (in 
$^{40}$Ca more $1\hbar\omega$ excitations are available).
Therefore, whenever the model space includes the whole $0\hbar\omega$
oscillator shell, a similar quenching is to be expected. The small
variations depend on the mass of the closed-shell core.

\begin{figure}
{\centering
\mbox{\psfig{figure=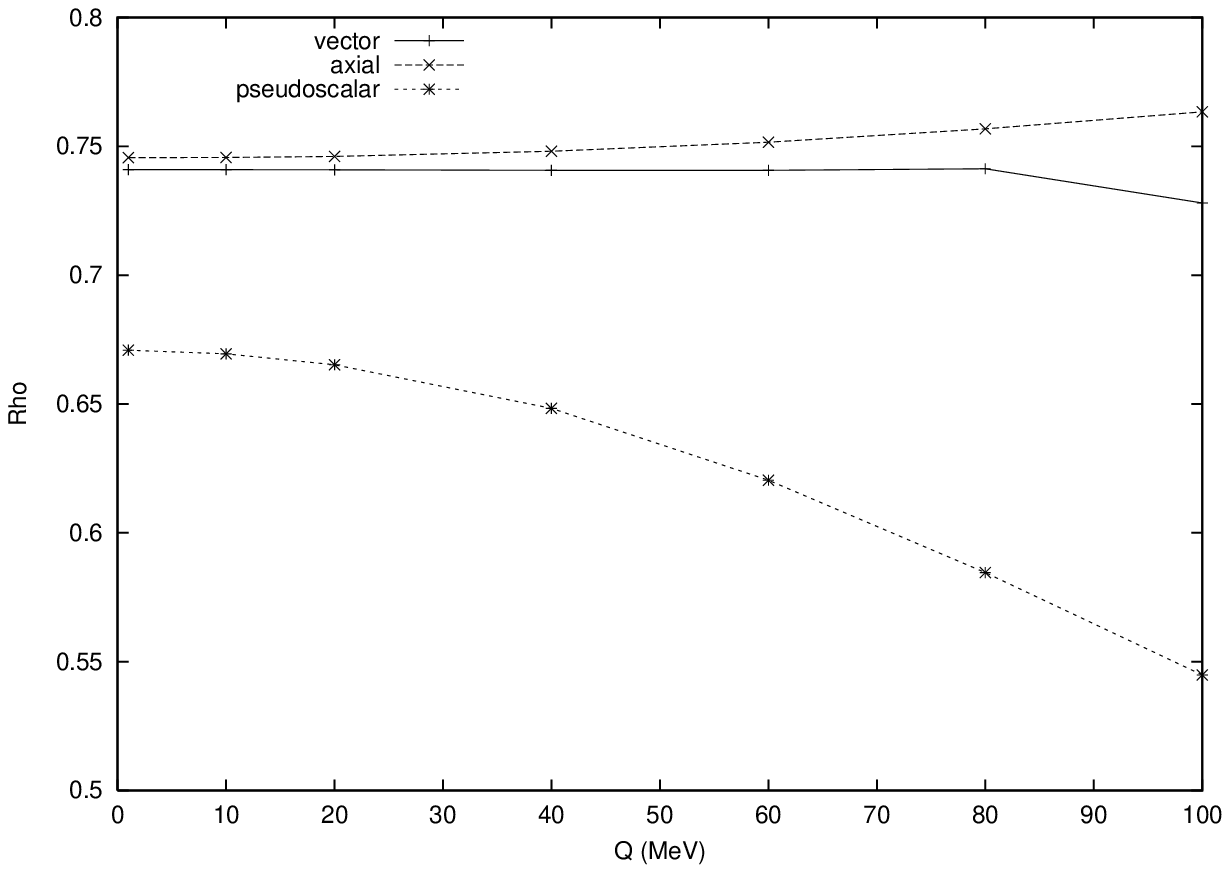,height=10cm,width=15cm,angle=0}}
}
  \caption{Renormalization of the vector, axial vector 
    and pseudoscalar terms with $^{56}$Ni as the closed-shell
    core, $\kappa=-1$ ($J=0\to 1$ transition, $\Delta\pi=\mathrm{no}$).
    \label{Fig:56}}
\end{figure}

The model space in which the effective operators are calculated for
$^{56}_{28}$Ni does not have a
closed $LS$ core. Now, the first-order transitions between $f_{7/2}$,
$f_{5/2}$, $g_{9/2}$, and $g_{7/2}$ single-particle orbits become
possible (the situation is
analogous to the $M1$ operator, which is diagonal in orbital angular
momentum and spin). This is clearly reflected in the values shown
in Fig.\ \ref{Fig:56}. We now have $\rho^2_A(0)=0.56$. We also remind
the reader that the Ikeda sum rule for Gamow--Teller beta decays
is not fulfilled in this space.

\begin{figure}
{\centering
\mbox{\psfig{figure=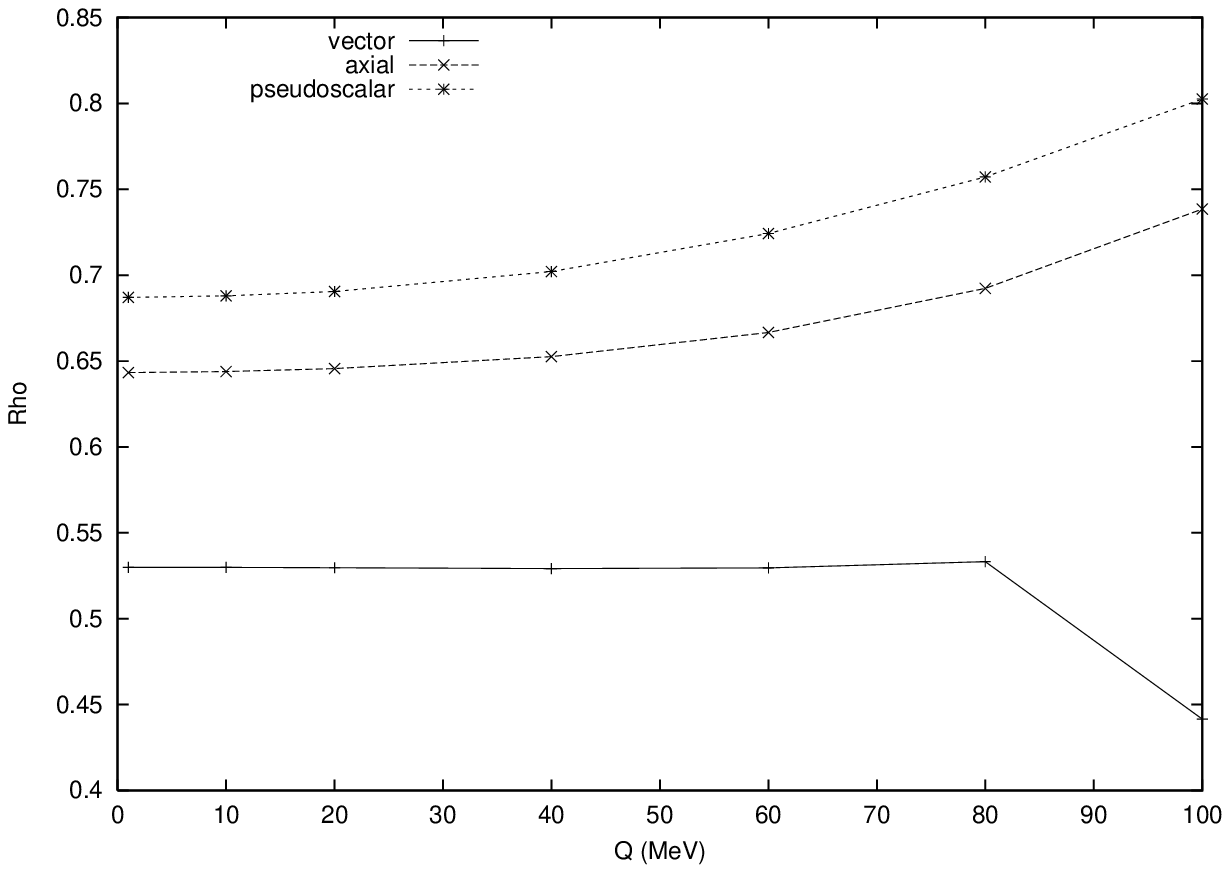,height=10cm,width=15cm,angle=0}}
}
  \caption{Renormalization of the vector, axial vector 
    and pseudoscalar terms with $^{100}$Sn as the closed-shell,
    $\kappa=-1$ ($J=0\to 1$ transition, $\Delta\pi=\mathrm{no}$).
    \label{Fig:100}}
\end{figure}

At mass $A=100$ our
space includes the single-particle orbits $0g_{7/2}$, $1d_{5/2}$, $1d_{3/2}$,
$2s_{1/2}$, and $0h_{11/2}$ above the $N=Z=50$ ($^{100}_{\phantom{1}50}$Sn)
major shell closure. The spin-orbit partners of $0g_{7/2}$ and 
$0h_{11/2}$ orbits, $0g_{9/2}$ and $0h_{9/2}$, are missing from our model
space. As in $^{56}$Ni, this is seen from
Fig.\ \ref{Fig:100}: All terms are quenched by a factor which is clearly
larger than in lighter nuclei with closed $LS$ shells. In the axial
vector part we have now $\rho^2_A(0)=0.41$, which
again stays nearly constant up to tens of MeV's, well beyond the beta
decay energy range (see also the discussion in Ref.\ \cite{kar98}).  
The diagonal transitions $g_{7/2}\to g_{7/2}$ and
$h_{11/2}\to h_{11/2}$ are responsible for the kink between $q=80$ and
100 MeV.

At the single-particle transition level, the spin-flip matrix elements (e.g.,
$f_{7/2}\to f_{5/2}$) are evidently more quenched than the diagonal 
ones. This feature does not depend on the operator or mass. We further note 
that the Gamow--Teller type matrix element $[101]$, being by
far the dominant axial term, follows very closely the trend of the axial
vector part through the whole mass range. 

\begin{figure}
{\centering
\mbox{\psfig{figure=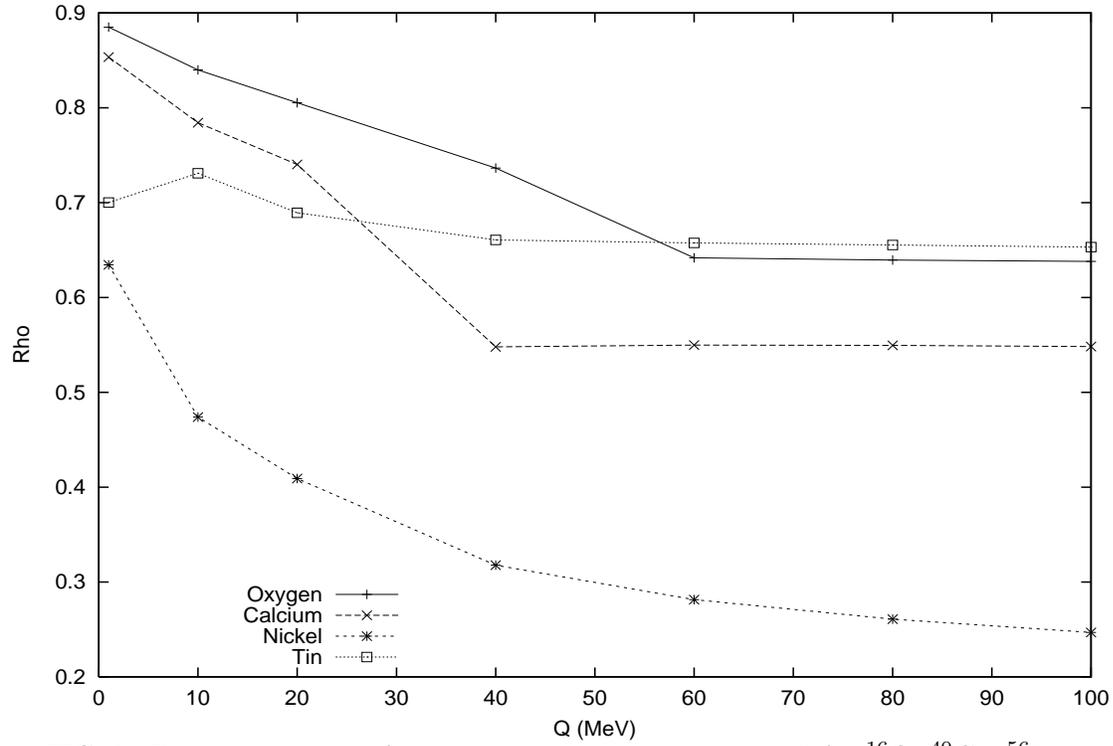,height=10cm,width=15cm,angle=0}}
}
  \caption{Renormalization of the axial vector term with
    $\kappa=2$ for $^{16}$O, $^{40}$Ca, $^{56}$Ni, and $^{100}$Sn 
    as closed-shell cores ($J=0\to 1$ transition, $\Delta\pi=\mathrm{no}$).
    \label{fig:kappa}}
\end{figure}

The $\kappa=2$ axial vector terms are shown in Fig.\ \ref{fig:kappa}. Only
the axial vector part is shown, since $V$ and $P$ terms are identical to
the case $\kappa=-1$. For $A=16$, 40, and 56 a strong quenching is obtained 
at high momentum transfers. This
is mainly caused by the spherical Bessel function $j_w(qr)$. Oscillations
in $\rho_A(q)$ are a sign of the interference between $j_w(qr)$ and the 
radial single-particle wave function. Clearly, a coherent extraction of
the renormalization is not as feasible as in the $\kappa=-1$ case.

In the report by Ciechanowicz {\em et al.} \cite{cie98}
the meson exchange contribution to the muon capture matrix elements
was found to be very small, at least in capture by $^{28}$Si. This is,
however, in contrast with the results for $A=12$ nuclei of Ref.\
\cite{gui82}. The quenching of the spin matrix element,
essentially the Gamow--Teller matrix element, is expected to be
dominated by the core polarization correction \cite{tow87}. 
However, our results leave some room for the subnucleonic
corrections. For example, in the $sd$ shell, about 50\% of the observed quenching
\cite{bro88} comes from the $\Delta$ isobars, meson exchange currents and
more complicated many-body terms. The situation is similar also in the 
beginning of the $fp$ shell ($^{40}$Ca).

The second order terms in the invariant amplitude describing the process
(\ref{reaction}) are proportional to $M^{-2}$. Therefore one would expect
their contribution to be at most a few percent of the first order terms
[see Eqs.\ (\ref{vector}), (\ref{axial}), and (\ref{pseudo})]. Indeed this
is the case. For example in $^{16}$O, the $r$-dependent single-particle 
matrix elements which,
according to Barabanov \cite{bar98}, should be the dominating second order
terms are about order of magnitude smaller than the first order terms. Only 
seldomly are  the magnitudes roughly equal (one case in $^{16}$O). Thus,
when the first-order terms are scaled by $M^{-1}$ and second-order terms by
$M^{-2}$, the magnitudes behave roughly as $50:1$, respectively.
Their contribution is, in this context,
neglible. The same conclusion is reached in \cite{bar98}, where detailed
expressions for the matrix elements are given. 

The quenching factors $\rho_A$ and $\rho_P$ can be used to estimate
the ratio $C_P/C_A$. We take the data from Figs.\ \ref{Fig:16},
\ref{Fig:40}, \ref{Fig:56}, and \ref{Fig:100} at $q=100$ MeV, which 
corresponds to
the muon capture region. Then we have $-3.5$\%, $-7.1$\%, $-28.6$\%,
and $+8.7$\%
changes in $C_P/C_A$ for masses 16, 40, 56, and 100, respectively. If the
bare value is taken from PCAC, $C_P/C_A\approx 8.4$, we have
$C_P/C_A\approx 8.1$, 7.8, 6.0, and 9.1 for masses 16, 40, 56, and
100. This yields an average $\sim 7.8$. 
 Although not directly comparable to our results, it is
interesting to note the results of Kolbe, Langanke, and Vogel 
\cite{kol94}. They used
continuum random phase approximation to calculate the part of the
capture rate which goes above the particle emission treshold. Their
results show a reasonable agreement with data when the bare couplings
are used. This nicely demonstrates the fact that if the model space
dimension is increased, the couplings should  asymptotically reach the
bare values.

\section{Summary}

We have constructed the effective transition operators corresponding to
the general form of the weak hadronic current between the proton and
neutron states. The effects of the renormalization are investigated as a
function of the transition $q$-value, and an average over the 
single-particle transitions is taken
separately for vector, axial vector and pseudoscalar terms. We have considered
only nucleonic degrees of freedom. In addition to the operators present
in the allowed beta decay, we have considered the higher-order corrections
to the transition amplitudes.

In the $sd$ and $fp$ shells, we get 19\% and 23\% quenchings in the axial 
vector strength, respectively. From these numbers we can conclude that
we have reached the large-$A$ limit already in the $fp$ shell, supporting the 
conclusions of \cite{mar96}. We have also explained this saturation in
qualitative terms. In $^{56}$Ni and $^{100}$Sn, where a major shell closure
separates the spin-orbit partners, a larger effect is seen. This is caused
by the first order contributions to the effective operator. The quenching
stays nearly constant for energies up to some 60 MeV in all cases. Therefore
it is justified to speak about energy-independent quenching factors for beta
decays in a given mass region. We also found that the
second-order terms (in inverse nucleon mass) are relatively unimportant for
most calculations. In particular, the uncertainties in the nuclear
model calculations of the one-body transition densities mask these tiny
corrections, generally a few percent at maximum.

The quenching factors are used to extract the value of the ratio $C_P/C_A$
at $q=100$ MeV. In light systems $^{16}$O and $^{40}$Ca our results indicate 
a small (of the order of few percent) quenching. In $^{100}$Sn, we obtain 
an enhancement of the same order of magnitude. In $^{56}$Ni a large
quenching is seen.

The next step in studies of effective operators
is the inclusion of subnucleonic degrees of freedom in the evaluation
of the different diagrams entering the definition of the effective
operator. Especially we have in mind the   
$\Delta$-isobars as an intermediate
state. These states have essentially been neglected due 
the lack of a suitable $\Delta\Delta$ interaction. 
We plan to extend our formalism to include such states through the use
of a newly refitted nucleon-nucleon interaction which includes
isobars as explicit degrees of freedom. This interaction \cite{mach2000} 
accounts for scattering data up to $\sim 1$ GeV in laboratory energy.  
The inclusion of meson exchange
effects together with the effective transition operators is also a
considerable
task, not fully attacked yet.

This work has been supported by the Academy of Finland under the
Finnish Centre of Excellence Programme 2000-2005 (Project No.\
44875, Nuclear and Condensed Matter Programme at JYFL).

\begin{table}
\caption{Reduced nuclear matrix elements and the
corresponding single-particle operators [without the lepton radial
wave function $G_\mu(r)$]. The 
$j_w(qr)$ are the spherical Bessel functions and $Y_{kwu}^M$ are the 
vector spherical harmonics~\protect\cite{mor60}. The momentum
operator for nucleons is $\bf p$, and $p_\nu$ is the momentum
of the neutrino.}
\begin{tabular}{ll}
        Matrix element & $O_{kwu}$ \\
        \hline
        \vspace{2pt}
        $[0wu]$ & $j_w(qr)Y_{0wu}^M(\hat {\bf r})\delta_{wu}$ \\
        $[1wu]$ & $j_w(qr)Y_{1wu}^M(\hat {\bf r},{\mathbf\sigma})$ \\
        \hline
        & $O_{kwu\pm}$ \\
        \hline
        $[0wu\pm]$ & $\left[j_w(qr)\pm\alpha Z(m_\mu'/p_\nu
           )j_{w\mp1}(qr)\right]Y_{0wu}^M
          (\hat {\bf r})\delta_{wu}$ \\
        $[1wu\pm]$ & $\left[j_w(qr)\pm\alpha Z(m_\mu'/p_\nu
           )j_{w\mp1}(qr)\right]Y_{1wu}^M
           (\hat {\bf r},{\mathbf\sigma})$ \\
        \hline
        & $O_{kwup}$ \\
        \hline
        $[0wup]$ & $ij_w(qr)Y_{0wu}^M(\hat
           {\bf r}){\mathbf\sigma}\cdot{\bf p}\delta_{wu}$ \\
        $[1wup]$ & $ij_w(qr)Y_{1wu}^M(\hat {\bf r},{\bf p})$ \\
\end{tabular}
\label{table:operators}
\end{table}

\clearpage
\clearpage
\clearpage
\clearpage
\clearpage

\end{document}